\documentclass{article}

\usepackage{amsmath,amssymb}
\usepackage{graphicx}
\usepackage{color}

\newtheorem{theorem}{Theorem}
\newtheorem{lemma}{Lemma}

\newtheorem{definition}{Definition}

\newtheorem{remark}{Remark}

\newcommand{\abu}{{\bf{a}}}

\newcommand{\bbu}{{\bf{b}}}
\newcommand{\vbu}{{\bf{v}}}
\newcommand{\ubu}{{\bf{u}}}
\newcommand{\dbu}{{\bf{d}}}
\newcommand{\tbu}{{\bf{t}}}
\newcommand{\cbu}{{\bf{c}}}

\newcommand{\sbu}{{\bf{s}}}
\newcommand{\wbu}{{\bf{w}}}

\newcommand{\lbu}{{\bf{l}}}

\newcommand{\ls}[1]
    {\dimen0=\fontdimen6\the\font\lineskip=#1\dimen0
     \advance\lineskip.5\fontdimen5\the\font
     \advance\lineskip-\dimen0
     \lineskiplimit=0.9\lineskip
     \baselineskip=\lineskip
     \advance\baselineskip\dimen0
     \normallineskip\lineskip\normallineskiplimit\lineskiplimit
     \normalbaselineskip\baselineskip
     \ignorespaces}

\begin{document}

\bibliographystyle{abbrv}

\title{Some Notes on Constructions  of  Binary  Sequences with Optimal Autocorrelation }
\author{Tongjiang Yan\\College of Science\\China University of Petroleum\\ Qingdao 266580, China\\
Email. yantoji@163.com\\
Guang Gong\\
Department of Electrical and Computer Engineering \\
University of Waterloo \\
Waterloo, Ontario N2L 3G1, CANADA \\
Email. ggong@calliope.uwaterloo.ca\\
}

\date{}
 \maketitle

\thispagestyle{plain}
\setcounter{page}{1}

\begin{abstract}
Constructions of binary sequences with low autocorrelation are considered in the paper. Based on recent progresses about this topic,
several more general constructions of binary sequences with optimal autocorrelations and other low autocorrelations are presented.


\end{abstract}

\ls{1.5}
\section{Introduction}
Pseudo random sequences with low cross correlation have important
applications in code-division multiple-access (CDMA)
communications and cryptology. The pseudo random sequences employed in CDMA communications
with low cross correlation may be
separated from the others in the family and
can successfully combat interference from the other users who
share a common channel. On the other hand, the sequences with
low cross correlation employed in either stream cipher cryptosystems
as key stream generators or in digital signature algorithms
as pseudo random number generators can resist correlation
attacks.

Given two binary sequences $a=(a(t))$ and $b=(b(t))$ of period $N$, the periodic correlation between  $a$ and $b$ is defined by
\begin{eqnarray}
R_{\abu\bbu}(\tau)=\sum_{t=0}^{N-1}(-1)^{a(t)+b(t+\tau)}, 0\leq \tau < N
\end{eqnarray}
where the addition $t+\tau$ is performed modulo $N$. Define the symbol $L^{m}(\abu)=\{a_{i+m}\}$. Then we have

\begin{lemma} \label{cross}  Let $m$ be an integer. Correlation of sequences satisfies the following properties:
\begin{eqnarray*}
&(1)& R_{L^{m}(\abu)\bbu}(\tau)=R_{\abu\bbu}(\tau-m),\\
&(2)& R_{\abu L^{m}(\bbu)}(\tau)=R_{\abu\bbu}(\tau+m), \\
&(3)& R_{\abu\bbu}(\tau)=R_{\abu\bbu}(\tau+N)=R_{\bbu\abu}(N-\tau),\\
&(4)& R_{\abu\bbu}(\tau)+R_{a\bar{\bbu}}(\tau)=R_{\abu\bbu}(\tau)+R_{\bar{\abu}\bbu}(\tau)=0.
\end{eqnarray*}
\end{lemma}

If $\abu=\bbu$, $R_{\abu\bbu}(\tau)$
is called the (periodic) autocorrelation function of $\abu$, denoted by $R_{\abu}(\tau)$, or simply $R(\tau)$ if the context is clear, otherwise, $R_{\abu\bbu}(\tau)$
is called the (periodic) cross-correlation function of $a$ and $b$.
For the autocorrelation of the sequence $a$, we have
\begin{lemma} \label{le1-a}
 Let $m$ be an integer. $R_{L^{m}(\abu)}(\tau)=R_{\bar{\abu}}(\tau)=R_{\abu}(\tau)$.
\end{lemma}

Let $\sbu=\{s_i\}$ denote a binary sequence of period $N$, The set
$C_{\sbu}=\{0\leq t\leq N-1:s(t)=1\}$
is called the support of $s$. If $k=\mid C_{\sbu}\mid$, then the periodic autocorrelation of $s(t)$ can be given by
\begin{eqnarray}\label{rakc}
R_{\sbu}(\tau) = N-4(k-\mid (\tau+C_{\sbu})\cap C_{\sbu}\mid),
\end{eqnarray}
where $k=\mid C_{\sbu}\mid$.
The smallest  possible  values for  the out-of-phase   autocorrelation function of a binary sequence  are listed below in Table \ref{tab1} depending on value of $N$ modulo 4 \cite{ARASU,cai}.  An autocorrelation function with one of  those values  is called the {\em perfect autocorrelation}.

\begin{table}
\caption{Perfect Autocorrelation Values for $\tau\not\equiv0 \bmod N$ }\label{tab1}
\begin{center}
\begin{tabular}{|| c | c  | p{2in}||}
\hline
 $N=i\,(\,\bmod{\,4})$ &  $R(\tau)$ & Comments\\
 \hline
0 & \{0\} & perfect sequence, only exists for $N = 4$, 0111, searched for $N < 108900$ \\\hline
1 & $\{1\}$ &  corresponding to $(2u^2+2u+1; u^2; u(u-1)/2)$ cyclic difference sets, exist for $u=1$ and $2$, not exist for $3\leq u<100$.\\ \hline
2 & $\{2\}$
or $ \{-2\}$ &  $R(\tau)=2$ does not exist for $N$ between 7 and 12545; $R(\tau)=-2$ only exists as the sequence $01$ or $10$. \\
\hline
3 & $-1$ & idea 2-level autocorrelation sequences, corresponding to cyclic Hadamard difference sets.  \\\hline
\end{tabular}
\end{center}
\end{table}

 The next  smallest  values for   the  out-of-phase autocorrelation of a binary sequence of period $N$ is listed below in Table \ref{tab2} \cite{ARASU,cai}, a  sequence with one of those autocorrelation is  called a sequence with {\em optimal autocorrelation}.

 \begin{table}
\caption{Optimal Autocorrelation Values  for $\tau\not\equiv0 \bmod N$}\label{tab2}
\begin{center}
\begin{tabular}{|| c | c  | p{2in}||}
\hline
 $N = i\,(\bmod{\,4})$ &  $R(\tau)$ & Comments\\
 \hline
0 & $\{0, -4\}$ & Sidelnikov sequences of period $q-1$, $q\equiv 1\bmod4$, Arasu-Ding-Helleseth-Kumar-Martinsen sequences, and some interleaved sequences in Table \ref{tab3}\\\hline
1 & $\{1, -3 \}$ & Legendre sequences of period $p\equiv1\bmod 4$, Ding-Helleseth-Lam sequences of period $p$, $p=x^2+4$ and $p\equiv1\bmod 4$, generalized cyclotomic sequences of period $p(p+4)$ \\ \hline
2 & $\{2, -2 \}$ & Sidelnikov sequences of period $q-1$, $q\equiv 3\bmod4$, Ding-Helleseth-Martinsen sequences\\ \hline
3 & $\{-1, 3\}$ &  Cai-Ding sequences \\\hline
\end{tabular}
\end{center}
\end{table}
For more details about ideal sequences and optimal autocorrelation, the reader is referred  to \cite{ARASU,cai,kd,dt,gg}.

Let $\tbu=(t(0),t(1),t(2),t(3))$ be a  binary sequence of period $4$, and the sequence
\begin{equation}\label{eq1}
\ubu = I(\abu_0+t(0), \abu_1+t(1), \abu_2+t(2), \abu_3+t(3))
\end{equation}
be an interleaved with $\abu_i+t(i)$ as its column sequences. Thus $\ubu$ has period $4N$ \cite{g}. Some known binary sequences with optimal autocorrelation are listed in the following Table \ref{tab3}, where  $L$ denotes the left shift operator and $H(\tbu)$ the hamming weight of the sequence $\tbu$.

\begin{table}
\caption{Progressive process  for finding  $(\abu_0, \abu_1, \abu_2, \abu_3)$ for being equal to $(\bbu_0,  L^{1/4+s}(\bbu_1), L^{1/2+s}(\bbu_2), L^{3/4+s}(\bbu_3))$ }\label{tab3}
\begin{center}
\begin{tabular}{|| c | c  | p{2in}||}
\hline
 $(\bbu_0, \bbu_1, \bbu_2, \bbu_3) $   &  $R_{\ubu}(\tau)$ & Comments\\
 \hline
 $(\abu, \abu, \abu, \abu)$ & $\{0, -4\}$ & $\abu$ 2-level auto Arash {\em et al.} 2001 $\tbu = 0111$  \cite{kd}. This form obtained by Yul and Gong 2008, and  $s = 0$, product sequence \cite{ng}. \\ \hline  \hline

  $(\abu', \abu', \abu, \abu)$ & $\{0, \pm 4\}$ & $\abu$ and $\abu'$ are paired $m$-sequences, $s = 0$ and $t_0+t_1 = 1$ by Yul and Gong, 2008, \cite{ng}\\ \hline
  \multicolumn{3}{||c||}{ The following cases are due to Tang and Gong, 2010, \cite{tg}}\\ \hline
 $(\abu', \abu', \abu, \abu)$ & $\{0, \pm 4\}$ & $\abu$ and $\abu'$ are paired GGMW or twin prime
paired
sequences.\\    \hline
 \multicolumn{3}{||c||}{ $\abu$ and $\abu'$ are paired Legendre sequences \cite{tg},} \\  \hline
  $(\abu, \abu', \abu, \abu')$  &  $\{0, - 4\}$ & $t_0 = 0$, $(t_1, t_2, t_3)\in \{001, 111\}$, $s = 0$ known before\\
 $(\abu, \abu, \abu', \abu')$  &  $\{0, \pm 4\}$ &  $t_0 = 0$, $H(t_1, t_2, t_3) = 1$ or 3\\

 $(\abu, \abu, \abu, \abu')$  &  $\{0, \pm 4\}$  & $t_0 = 0$, $H(t(1), t(2), t(3)) = 1$ or 3\\

 \hline \hline

 $(\abu, \bbu,  \abu, \bbu)$  & $\{0, -4\}$ & $\abu$ and $\bbu$ are 2-level, $s = 0$, and $\tbu = 0001$, Tang and Ding, 2010 \cite{tg}. Equivalently, it is also true for $\tbu = 0010, 1101$ or $1110$.\\

 \hline

\end{tabular}
\end{center}
\end{table}

This paper contributes to give a  general construction different from the one in \cite{tg}, which is a generalization of the construction in \cite{Tangding}. By inputting some perfect sequences and three pairs of  sequences in \cite{tg} into our new construction respectively, several kinds of sequences with optimal autocorrelation or other low autocorrelation can be produced.

\section{Three interleaved sequences and their modifications }

Let $\sbu=(s(0), s(1), \cdots, s(N-1))$ be a binary sequence of period $N$ and can be denoted by the following interleaved constructions:

Construction A: $\sbu=I(\mathbf{0} _{K}, \abu_1, \abu_2, \cdots, \abu_{T-1})$ is a generalized GMW construction defined in \cite{tg},
where
\begin{itemize}
\item(1) $\mathbf{0}_{K}$ is an all zero sequence of period $K$.
\item(2) $\abu_i, 1\leq i\leq T-1, $ are shift equivalent and possess ideal autocorrelation.
\end{itemize}

By replacing the first column sequence $\mathbf{0}_{K}$ in the sequence $s$ with $\mathbf{1}_{K}$, we get the following modified construction of  Construction A:

Construction B: $s'=I(\mathbf{1}_{K}, \abu_1, \abu_2, \cdots, \abu_{T-1})$.

Define
$
d(a_{i})=\mid \{k\mid a_{i}(k)=1, k=0, 1, 2, \cdots, K-1\}\mid
-\mid \{k\mid a_{i}(k)=0, k=0, 1, 2, \cdots K-1\}\mid.
$
Then we have obtained the following results:
\begin{lemma}\cite{yan}\label{s2} For the sequences $s$ and $s'$,
\begin{eqnarray*}
R_{s's}(\tau)&=&R_{s s'}(\tau)\Longleftrightarrow d(a_{T-\tau_2})= d(a_{\tau_2}).
\end{eqnarray*}
\end{lemma}
\begin{lemma} \cite{tg}\label{gmw} Let $\sbu$ and $\sbu'$ be the sequences defined above, $T=2^n+1,K=2^n-1$.
\begin{eqnarray*}
R_\sbu(\tau)&=&\left\{
\begin{array}{lll}
2^{2n}-1   &\mbox{if } \tau=0, \\
-1&\mbox{ otherwise }.
\end{array}
\right.\\
R_{\sbu'}(\tau)&=&\left\{
\begin{array}{lll}
2^{2n}-1   &\mbox{if } \tau=0, \\
-1   &\mbox{if }  \tau\equiv0 \bmod 2^n+1,\tau\not=0, \\
3  &\mbox{ otherwise }.
\end{array}
\right.\\
R_{\sbu'\sbu}(\tau)&=&R_{\sbu\sbu'}(\tau)=\left\{
\begin{array}{llll}
2^{2n}-2^{n+1}+1   &\mbox{ if } \tau=0, \\
-2^{n+1}+1   &\mbox{ if }  \tau\equiv0 \bmod 2^n+1,\tau\not=0, \\
1  &\mbox{ otherwise }.
\end{array}
\right.
\end{eqnarray*}
\end{lemma}

Let $p$ be a odd prime and  Legendre function
\begin{eqnarray*}(\frac{t}{p})=\left\{
\begin{array}{lll}
1  & \mbox{ if }  t\in QR_p, \\
-1 &\mbox{ if }  t\in NQR_p.
\end{array}
\right.
\end{eqnarray*}
where $QR_p$ and $NQR_p$ denote the quadratic residue and nonquadratic residue of $p$.

a Legendre sequence $l(t)$ is defined as
\begin{eqnarray*}l(t)=\left\{
\begin{array}{lll}
0\mbox{ or }1 &\mbox{if } t=0, \\
\frac{1}{2}(1-(\frac{t}{p}))&\mbox{otherwise}.
\end{array}
\right.
\end{eqnarray*}
$l(t)$ is called the first type Legendre sequence  if $l(0)=1$ otherwise the second type Legendre sequence (denoted by $l'(t)$).

\begin{lemma} \cite{tg} \label{R-legendre} Legendre sequences $l(t)$ and $l'(t)$ possess the following autocorrelation.

If $p\equiv3\bmod 4$, $l(t)$ and $l'(t)$ possess ideal autocorrelation.

If $p\equiv1\bmod 4$,
\begin{eqnarray*}R_\lbu(\tau)&=&\left\{
\begin{array}{lll}
p    &\mbox{if } \tau=0, \\
1   &\mbox{if }  \tau\in QR_p, \\
-3  &\mbox{if }  \tau\in NQR_p.
\end{array}
\right.\\
R_{\lbu'}(\tau)&=&\left\{
\begin{array}{lll}
p    &\mbox{if } \tau=0, \\
-3   &\mbox{if } \tau \in QR_p, \\
1   &\mbox{if } \tau \in NQR_p.
\end{array}
\right.
\end{eqnarray*}
\end{lemma}

\begin{lemma} \cite{tg} \label{Rll-legendre} Legendre sequences $l(t)$ and $l'(t)$ possess the following crosscorrelation.

If $N\equiv 1\bmod4$,
\begin{eqnarray*}
R_{\lbu\lbu'}(\tau)&=&R_{\lbu'\lbu}(\tau)=\left\{
\begin{array}{lll}
N-2   &\mbox{if } \tau=0, \\
-1  &\mbox{ otherwise }.
\end{array}
\right.
\end{eqnarray*}

If $N\equiv 3\bmod4$,
\begin{eqnarray*}
R_{\lbu\lbu'}(\tau)&=\left\{
\begin{array}{lll}
N-2   &\mbox{if } \tau=0, \\
1   &\mbox{if }  \tau\in QR_p, \\
-3  &\mbox{if }  \tau\in NQR_p.
\end{array}
\right.\\
R_{\lbu'\lbu}&=\left\{
\begin{array}{lll}
N-2   &\mbox{if } \tau=0, \\
-3   &\mbox{if } \tau\in QR_p, \\
1  &\mbox{if }  \tau\in NQR_p.
\end{array}
\right.
\end{eqnarray*}
\end{lemma}

For the twin-prime sequence
$$\tbu=I(\mathbf{0}_p, L^{e_1}(a_1)+b(1), \cdots L^{e_{p+1}}(a_{p+1})+b(p+1))$$
where $e_i=i(p+2)^{-1}\bmod p$, $p$ and $p+2$ are two primes, $b(i)=1$ if $i\in QR_{p+2}$ otherwise $b(i)=0$, and $a_i=l'$ if $i\in QR_{p+2}$ otherwise $a_i=l, i=1, 2, \cdots, p+1$.The modified type of the twin-prime sequence $t$
\begin{eqnarray*}
\tbu'=I(\mathbf{1}_p, L^{e_1})(a_1)+b(1), \cdots L^{e_{p+1}})(a_{p+1})+b(p+1))
\end{eqnarray*}

\begin{lemma}\cite{tg} \label{twin}The twin-prime sequence and its modified possess the following correlation properties.
\begin{eqnarray*}
R_{\tbu}(\tau)&=&\left\{
\begin{array}{lll}
p(p+2)   &\mbox{if } \tau=0, \\
-1 &\mbox{ otherwise }.
\end{array}
\right.\\
R_{\tbu'}(\tau)&=&\left\{
\begin{array}{lll}
p(p+2)  &\mbox{if } \tau=0, \\
-1   &\mbox{if }  \tau\equiv 0\bmod (p+2),\tau\not=0, \\
3  &\mbox{otherwise }  .
\end{array}
\right.\\
R_{\tbu\tbu'}(\tau)&=&R_{\tbu'\tbu}(\tau)=\left\{
\begin{array}{lll}
p^2  &\mbox{if } \tau=0, \\
-2p-1  &\mbox{if }  \tau\equiv 0\bmod (p+2),\tau\not=0, \\
1  &\mbox{otherwise }  .
\end{array}
\right.
\end{eqnarray*}
\end{lemma}

\section{Correlation Properties of Sequences in Two Difference Constructions}

For two binary sequences $\abu=\{a_i\}$ and $\bbu=\{b_i\}$ of period $N$, a sequence  $\ubu=\{u_j\}\triangleq\{a_i\parallel b_i\}=\abu\parallel \bbu$ is defined as
$$u_j=\left\{
\begin{array}{lll}
a_i& \mbox{~if~} j=2i,\\
b_i& \mbox{~if~} j=2i+1.
\end{array}
\right.
$$

Define another two binary sequences $\cbu=\{c_i\}$ and $\dbu=\{d_i\}$ of period $N$.

\begin{lemma} \label{abcd} The correlation
\begin{eqnarray}
 R_{(\abu\parallel \bbu)(\cbu\parallel \dbu)}(\tau)=\left\{
\begin{array}{lll}
R_{\abu\cbu}(\frac{\tau}{2})+R_{\bbu\dbu}(\frac{\tau}{2})& \mbox{~if~} \tau \mbox{~is even},\\
R_{\abu\dbu}(\frac{\tau-1}{2})+R_{\bbu\cbu}(\frac{\tau+1}{2})& \mbox{~if~} \tau \mbox{~is odd}.
\end{array}
\right.
\end{eqnarray}
\end{lemma}

Assuming $\abu=\cbu$ and $\bbu=\dbu$ in the above Lemma \ref{abcd}, then we get
\begin{lemma}\label{apb}
The autocorrelation of $\ubu=\abu\parallel \bbu$
\begin{eqnarray}
R_\ubu(\tau)=\left\{
\begin{array}{lll}
R_\abu(\frac{\tau}{2})+R_\bbu(\frac{\tau}{2})& \mbox{~if~} \tau \mbox{~is even},\\
R_{\abu\bbu}(\frac{\tau-1}{2})+R_{\bbu\abu}(\frac{\tau+1}{2}) & \mbox{~if~} \tau \mbox{~is odd}.
\end{array}
\right.
\end{eqnarray}
\end{lemma}

Specially, let $\bbu=L^m(\abu)$ in the above Lemma \ref{apb}.
\begin{lemma} \label{aplma}  The autocorrelation of $\ubu=\abu\parallel L^m(\abu)$
\begin{eqnarray}
R_\ubu(\tau)=\left\{
\begin{array}{lll}
2R_\abu(\frac{\tau}{2})& \mbox{~if~} \tau \mbox{~is even},\\
R_{\abu}(\frac{\tau-1}{2}+m)+R_{\abu}(\frac{\tau+1}{2}-m) & \mbox{~if~} \tau \mbox{~is odd}.
\end{array}
\right.
\end{eqnarray}

If $N$ is odd and $m=\frac{N+1}{2}$, then the autocorrelation of $\ubu=\abu\parallel L^m(\abu)$
\begin{eqnarray}
R_\ubu(\tau)=\left\{
\begin{array}{lll}
2R_\abu(\frac{\tau}{2})& \mbox{~if~} \tau \mbox{~is even},\\
2R_{\abu}(\frac{\tau+N}{2})& \mbox{~if~} \tau \mbox{~is odd}.
\end{array}
\right.
\end{eqnarray}
\end{lemma}

Specially, let $\bbu=L^m(\bar{\abu})$  in the above Lemma \ref{apb}.
\begin{lemma} \label{aplma8} The autocorrelation of $\ubu=\abu\parallel L^m(\bar{\abu})$
\begin{eqnarray}
R_\ubu(\tau)=\left\{
\begin{array}{lll}
2R_\abu(\frac{\tau}{2})& \mbox{~if~} \tau \mbox{~is even},\\
-R_{\abu}(\frac{\tau-1}{2}+m)-R_{\abu}(\frac{\tau+1}{2}-m) & \mbox{~if~} \tau \mbox{~is odd}.
\end{array}
\right.
\end{eqnarray}

If $N$ is odd and $m=\frac{N+1}{2}$, then the autocorrelation of $\ubu=\abu\parallel L^m(\bar{\abu})$
\begin{eqnarray}
R_\ubu(\tau)=\left\{
\begin{array}{lll}
2R_\abu(\frac{\tau}{2})& \mbox{~if~} \tau \mbox{~is even},\\
-2R_{\abu}(\frac{\tau+N}{2})& \mbox{~if~} \tau \mbox{~is odd}.
\end{array}
\right.
\end{eqnarray}
\end{lemma}

\begin{lemma} \label{apla8bplb} The correlation
\begin{eqnarray}
 R_{(\abu\parallel L^{\frac{N+1}{2}}(\bar{\abu}))(\bbu\parallel L^{\frac{N+1}{2}}(\bbu))}(\tau)=R_{(\bbu\parallel L^{\frac{N+1}{2}}(\bbu)(\abu\parallel L^{\frac{N+1}{2}}(\bar{\abu}))}(\tau)=0.
\end{eqnarray}
\end{lemma}

\noindent\textbf{Proof. } If $\tau$ is even, then, from Lemmas \ref{abcd} and \ref{cross},
\begin{eqnarray} \label{eq1-1}R_{(\abu\parallel L^{\frac{N+1}{2}}(\bar{\abu}))(\bbu\parallel L^{\frac{N+1}{2}}(\bbu)}(\tau)
&=&R_{\abu\bbu}(\frac{\tau}{2})+R_{L^{\frac{N+1}{2}}(\bar{\abu})L^{\frac{N+1}{2}}(\bbu)}(\frac{\tau}{2})\\
&=&R_{\abu\bbu}(\frac{\tau}{2})+R_{\bar{\abu}\bbu}(\frac{\tau}{2})=0.\nonumber
\end{eqnarray}
If $\tau$ is odd, then, from Lemmas \ref{abcd} and \ref{cross},

\begin{eqnarray}\label{eq1-2}R_{(\abu\parallel L^{\frac{N+1}{2}}(\bar{\abu}))(\bbu\parallel L^{\frac{N+1}{2}}(\bbu)}(\tau)
&=&R_{\abu L^{\frac{N+1}{2}}(\bbu)}(\frac{\tau-1}{2})+R_{L^{\frac{N+1}{2}}(\bar{\abu})\bbu}(\frac{\tau+1}{2})\\
&=&R_{\abu\bbu}(\frac{\tau-1}{2}+\frac{N+1}{2})+R_{\bar{\abu}\bbu}(\frac{\tau+1}{2}-\frac{N+1}{2})\nonumber\\
&=&R_{\abu\bbu}(\frac{\tau+N}{2})+R_{\bar{\abu}\bbu}(\frac{\tau-N}{2})\nonumber\\
&=&0.\nonumber
\end{eqnarray}
Thus, from Equations  (\ref{eq1-1}) and (\ref{eq1-2}),
\begin{eqnarray} \label{eq13}
R_{(\abu\parallel L^{\frac{N+1}{2}}(\bar{\abu}))(\bbu\parallel L^{\frac{N+1}{2}}(\bbu))}(\tau)=0.
\end{eqnarray}
And, from Lemma \ref{cross} and  Equation \ref{eq13},
\begin{eqnarray*}
R_{(b\parallel L^{\frac{N+1}{2}}(\bbu)(\abu\parallel L^{\frac{N+1}{2}}(\bar{\abu}))}(\tau)=R_{(\abu\parallel L^{\frac{N+1}{2}}(\bar{\abu}))(b\parallel L^{\frac{N+1}{2}}(\bbu))}(2N-\tau)=0.
\end{eqnarray*}

\begin{definition} \label{st} Define a new binary interleaved sequence $\vbu$ as the following
\begin{eqnarray*}
v(t) = (\abu, \bbu, L^{\frac{N+1}{2}}(\bar{\abu}), L^{\frac{N+1}{2}}(\bbu)).
\end{eqnarray*}
\end{definition}
where $\abu$ and $\bbu$ are two binary sequences with length $N, N\equiv 1,3\bmod4$.

\begin{theorem} \label{main} Let $\tau = 4\tau_1+\tau_2, 0\leq \tau_2\leq 3$. Autocorrelation of  the   sequence $\vbu$ is
\begin{eqnarray*}
R_{\vbu}(\tau) = \left\{
\begin{array}{ll}
 2R_{\abu}(\tau_1) +2R_{\bbu}(\tau_1) &\mbox{ if }\tau_2 = 0,\\
 0&\mbox{ if }  \tau_2 = 1, 3, \\\vspace{1mm}
 -2R_{\abu}(\tau_1+\displaystyle\frac{N+1}{2})+2R_{\bbu}(\tau_1+\displaystyle\frac{N+1}{2})&\mbox{ if } \tau_2 = 2.
\end{array}
\right.
\end{eqnarray*}
\end{theorem}

\noindent\textbf{Proof. }
Actually, the sequence $v$ can be seen as $(\abu\parallel L^{\frac{N+1}{2}}(\bar{\abu}))\parallel(b\parallel L^{\frac{N+1}{2}}(\bbu))$. Thus, from Lemma \ref{aplma},
\begin{eqnarray*}
R_{\vbu}(\tau) &=& \left\{
\begin{array}{ll}
 R_{\abu\parallel L^{\frac{N+1}{2}}(\bar{\abu})}(\frac{\tau}{2}) + R_{b\parallel L^{\frac{N+1}{2}}(\bbu)}(\frac{\tau}{2}) &\mbox{ if }\tau \mbox{ is even,}\\
 R_{(\abu\parallel L^{\frac{N+1}{2}}(\bar{\abu})(b\parallel L^{\frac{N+1}{2}}(\bbu))}(\frac{\tau-1}{2})+ R_{(b\parallel L^{\frac{N+1}{2}}(\bbu))(\abu\parallel L^{\frac{N+1}{2}}(\bar{\abu}))}(\frac{\tau+1}{2})  &\mbox{ if }\tau \mbox{ is odd. }
 \end{array}
\right.
\end{eqnarray*}

For the case $\tau$ is even, if  $\tau_2=0$, then $\frac{\tau}{2}$ is even, from Lemmas (\ref{apb}),(\ref{aplma}),(\ref{aplma8}), and (\ref{le1-a}),
\begin{eqnarray*}
R_{\vbu}(\tau)&=&R_{\abu}(\frac{\tau}{4})+R_{L^{\frac{N+1}{2}}(\bar{\abu})}(\frac{\tau}{4})+ R_{\bbu}(\frac{\tau}{4})+ R_{L^{\frac{N+1}{2}}(\bbu)}(\frac{\tau}{4})\\
           &=&2R_{\abu}(\tau_1)+ 2R_{\bbu}(\tau_1).
\end{eqnarray*}
if  $\tau_2=2$, then $\frac{\tau}{2}$ is odd, from Lemmas (\ref{abcd}),(\ref{aplma}) and (\ref{aplma8}),
\begin{eqnarray*}
R_{\vbu}(\tau)&=&-2R_{\abu}(\frac{4\tau_1+2+2N}{4})+2R_{\bbu}(\frac{4\tau_1+2+2N}{4})\\
           &=&-2R_{\abu}(\tau_1+\frac{N+1}{2})+ +2R_{\bbu}(\tau_1+\frac{N+1}{2}).
\end{eqnarray*}

For the case $\tau$ is odd, from Lemma \ref{apla8bplb},
$
R_{\vbu}(\tau)=0.
$

Based on Theorem \ref{main}, many binary sequences with low autocorrelation can be obtained as the following Theorems \ref{n34}$-$ \ref{tt}

\begin{theorem} \label{n34} If $N\equiv 3\bmod4$, then $\ubu(t)$ possess optimal autocorrelation $R_{\ubu}(\tau)\in \{4N, -4, 0\}$ if and only if  $\abu$ and $\bbu$ in Definition \ref{st} are two binary sequences with ideal autocorrelation.

At this time, the  autocorrelation function of $\ubu(t)$
\begin{eqnarray*}
&&R_{\ubu}(\tau) = \left\{
\begin{array}{ll}
 4N& \mbox{ if }  \tau = 0,\\
-4 &\mbox{ if } \tau_2 = 0\\
                & \mbox{ and } \tau\neq0,\\
0&\mbox{otherwisetimes}.
\end{array}
\right.
\end{eqnarray*}
\end{theorem}

\noindent\textbf{Proof. } Sufficiency can be verified directly by Theorem \ref{main}. For the necessity,  from Theorem \ref{main}, the statement $s$ possesses optimal autocorrelation $R_{\sbu}(\tau)\in \{4N, -4, 0\}$ requires that $R_{\abu}, R_{\bbu}\in \{0, \pm 2,\pm 1\}$. But, by Equation (\ref{rakc}), neither $R_{\abu}$ nor $R_{\bbu}$  of  period $N\equiv 3\bmod4$ can take values $0,1,\pm 2$. Thus there exists only one choice that $R_{\abu} = R_{\bbu} = -1$.
\begin{remark}
It is easy to prove the above Theorem \ref{n34} is equivalent to the Construction B in \cite{Tangding} and is also true in the cases $\tbu = 0001, 1101$ or $1110$. The corresponding almost different set has been given in Theorem 9 \cite{Tangding}.
\end{remark}

Similarly to the proof of the above Theorem \ref{n34}, we can prove the following Theorem \ref{n14no}:
\begin{theorem} \label{n14no}
If $N\equiv 1\bmod4$, $\ubu(t)$ possesses optimal autocorrelation $R_{\sbu}\in \{4N,4,0\}$ if and only if $\abu$ and $\bbu$ possess optimal autocorrelation $R_{\abu}, R_{\bbu}\in \{N,  1\}$. At this time, the  autocorrelation function of $\ubu(t)$
\begin{eqnarray*}
&&R_{\ubu}(\tau) = \left\{
\begin{array}{ll}
 4N& \mbox{ if }  \tau = 0, \\
4 &\mbox{ if } \tau_2 = 0\\
                & \mbox{ and } \tau\neq0,\\
0&\mbox{otherwise}.
\end{array}
\right.
\end{eqnarray*}
\end{theorem}

Let $\abu=01000,\bbu=10000$, then $\ubu(t)=01101010001100000010$ possesses optimal autocorrelation $R_{\sbu}\in \{4N,4,0\}$. If $\bbu=10000$ is replaced by its shift $\bbu'=00010$, then the corresponding $\ubu'(t)=00111010001001000010$ also possesses optimal autocorrelation in $\{4N,4,0\}$. Obviously, $\ubu(t)\not=\ubu'(t)$. But their supports are the same almost difference set $(20, 7 ,3, 2)$. Since there only exist two known binary sequences  possess optimal autocorrelation in $\{N,  1\}$, from which we can only get finite sequences with optimal autocorrelation in $\{4N,4,0\}$. But, as we know, these are only binary sequences with this type of optimal autocorrelation.

By Lemma \ref{gmw} and  Theorem \ref{main}, we have

\begin{theorem} \label{4444}
If $\sbu$ and $\sbu'$ are generalized GMW sequence Construction  A  and its modified type in Construction B, and $(\abu,\bbu)=(L^{\eta_1}(\sbu), L^{\eta_2}(\sbu'))$ $($or $ (L^{\eta_1}(\sbu'), L^{\eta_2}(\sbu)))$, where $0\leq \eta_1,\eta_2\leq 2^{2n}-2$, then $\ubu(t)$ possesses autocorrelation as the following
\begin{eqnarray*}
R_{\ubu}(\tau) = \left\{
\begin{array}{llll}
2^{2n+2}-4   &\mbox{ if } \tau = 0, \\
-4   &\mbox{ if }  \tau_1\equiv0 \bmod 2^n+1, \tau_2\equiv 0\bmod4, \tau\not = 0, \\
4   &\mbox{ if }   \tau_1\not\equiv0 \bmod 2^n+1, \tau_2\equiv 0\bmod4, \\
0  &\mbox{ if }    \tau_2\equiv 1,3\bmod4, \\
0  &\mbox{ if }   \tau_1+2^{2n-1}\equiv0 \bmod 2^n+1, \tau_2=2, \\
8 \,\,( or -8)   &\mbox{ if }   \tau_1+2^{2n-1}\not\equiv0 \bmod 2^n+1, \tau_2=2, \\
\end{array}
\right.
\end{eqnarray*}
\end{theorem}

Note that $\displaystyle\frac{1}{2} = 2^{2n-1}$ in  the ring $Z_{2^{2n}-1}$.

By Lemma \ref{R-legendre} and Theorem \ref{main}, we have
\begin{theorem} \label{804-4}Let $\mathbf{l}', \mathbf{l}$  be two types of Legendre sequence respectively of period $p$, $p\equiv 1 \bmod 4$,   and $(\abu,\bbu)=(L^{\eta_1}(\mathbf{l}'), L^{\eta_2}(\mathbf{l}))$ $($ or $((L^{\eta_1}(\mathbf{l}), L^{\eta_2}(\mathbf{l}'))))$, where $0\leq \eta_1,\eta_2\leq p-1$, then $\ubu(t)$ have autocorrelation as the following
\begin{eqnarray*}
&&R_{\ubu}(\tau) = \left\{
\begin{array}{lllll}
4p& \mbox{ if }  \tau = 0,\\
-4&\mbox{ if }   \tau\neq 0 ,\tau_2 = 0,\\
         0&\mbox{ if } \, \, \tau_2 = 1, 3,\\
0 &\mbox{ if } \tau_1+\displaystyle\frac{1}{2} = 0,\tau_2 = 2,\\
8 \,\,( or  -8)&\mbox{ if } 0\neq\tau_1+\displaystyle\frac{1}{2} \in QR_p, \\
                 & \mbox{ and }\, \,\tau_2 = 2,\\
-8 \,\,( or \,\, 8) &\mbox{ if }   0\neq\tau_1+\displaystyle\frac{1}{2} \in NQR_p,\\
  & \mbox{ and }\, \, \tau_2 = 2.\\
\end{array}
\right.
\end{eqnarray*}
\end{theorem}

By Lemma \ref {twin} and Theorem \ref{main}, we have
\begin{theorem} \label{tt} Let $\tbu$ and $\tbu'$ be the twin-prime sequence and its modification,  and $(\abu,\bbu)=(L^{\eta_1}(\tbu), L^{\eta_2}(\tbu'))$ $($or $((L^{\eta_1}(\tbu'), L^{\eta_2}(\tbu)))))$, where $0\leq \eta_1,\eta_2\leq p(p+2)-1$, then  the interleaved sequence $\ubu(t)$ have autocorrelation as the following
\begin{eqnarray*}
&&R_{\ubu}(\tau) = \left\{
\begin{array}{lllll}
4p(p+2)& \mbox{ if }  \tau = 0,\\
-4   &\mbox{ if }   \tau\neq 0 ,\tau_2 = 0, \tau_1 \equiv 0\bmod (p+2),\\
4   &\mbox{ if }    \tau_2 = 0, \tau_1\not\equiv 0\bmod (p+2),\\
0&\mbox{ if } \, \, \tau_2 = 1, 3, \\ \vspace{1mm}
0&\mbox{ if } \tau_1+\displaystyle\frac{1}{2} \equiv 0\bmod (p+2),\tau_2 = 2,\\
8 \,\,( or -8) &\mbox{ if }  \tau_1+\displaystyle\frac{1}{2} \not\equiv 0\bmod (p+2), \tau_2 = 2.\\
\end{array}
\right.
\end{eqnarray*}
\end{theorem}

\begin{definition} \label{st} Define a new binary interleaved sequence $\wbu$ as the following
\begin{eqnarray*}
w(t)=(\abu, L^{\eta}(\bar{\abu}), \bbu, L^{\eta}(\bbu)).
\end{eqnarray*}
\end{definition}
where $\abu$ and $\bbu$ are two binary sequences with length $N, N\equiv 1,3\bmod4$.

\begin{theorem} \label{main2} Let $\tau = 4\tau_1+\tau_2, 0\leq \tau_2\leq 3$. Autocorrelation of  the   sequence $\wbu$ is
\begin{eqnarray*}
R_{\wbu}(\tau) = \left\{
\begin{array}{ll}
 2R_{\abu}(\tau_1) +2R_{\bbu}(\tau_1) &\mbox{ if }\tau_2 = 0,\\
 -R_{\abu}(\tau_1+\eta)+ R_{\bbu}(\tau_1+\eta) &\mbox{ if } \tau_2 = 1 \\
 -R_{\abu\bbu}(\tau_1-\eta)
 +R_{\bbu \abu}(\tau_1-\eta)& \\
 0&\mbox{ if } \tau_2 = 2,\\
-R_{\abu}(\tau_1+1-\eta)+R_{\bbu}(\tau_1+1-\eta)&\mbox{ if } \tau_2=3.\\
 +R_{\abu\bbu}(\tau_1+\eta)-R_{\bbu\abu}(\tau_1+\eta)&.
\end{array}
\right.
\end{eqnarray*}
\end{theorem}

\noindent\textbf{Proof. }
The sequence $\wbu$ can be seen as $$(\abu\parallel \bbu)\parallel(L^{\eta}(\bar{\abu})\parallel L^{\eta}(\bbu))=(\abu\parallel \bbu)\parallel(L^{\eta}(\bar{\abu}\parallel b)).$$ Thus, from Lemma \ref{aplma},
\begin{eqnarray}\label{eqRw}
R_{w}(\tau) &=& \left\{
\begin{array}{ll}
 R_{\abu\parallel \bbu}(\frac{\tau}{2}) + R_{L^{\eta}(\bar{\abu}\parallel b)}(\frac{\tau}{2}) &\mbox{ if } \tau \mbox{ is even,}\\
 R_{(\abu\parallel \bbu)(L^{\eta}(\bar{\abu}\parallel b))}(\frac{\tau-1}{2})+ R_{(L^{\eta}(\bar{\abu}\parallel b))(\abu\parallel \bbu)}(\frac{\tau+1}{2})  &\mbox{ if }\tau \mbox{ is odd. }
 \end{array}
\right.
\end{eqnarray}

For the case $\tau$ is even, if  $\tau_2=0$, then $\frac{\tau}{2}$ is even, from Lemmas \ref{apb} and \ref{le1-a} and Equation (\ref{eqRw}),
\begin{eqnarray*}
R_{w}(\tau)&=&R_{\abu}(\frac{\tau}{4})+ R_{\bbu}(\frac{\tau}{4})+ R_{\bar{\abu}}(\frac{\tau}{4})+R_{\bbu}(\frac{\tau}{4})\\
           &=&2R_{\abu}(\tau_1)+ 2R_{\bbu}(\tau_1).
\end{eqnarray*}
if  $\tau_2=2$, then $\frac{\tau}{2}$ is odd, from Lemmas \ref{le1-a}, \ref{cross} and Equation (\ref{eqRw}),
\begin{eqnarray*}
R_{w}(\tau)&=&R_{\abu\bbu}(\frac{\tau-2}{4})+R_{ba}(\frac{\tau+2}{4})+R_{\bar{\abu}b}(\frac{\tau-2}{4})+R_{b\bar{\abu}}(\frac{\tau+2}{4})\\
           &=&0.
\end{eqnarray*}

For the case $\tau$ is odd, if $\tau_2=1$, then $\frac{\tau-1}{2}$ is even, from Lemmas \ref{abcd},\ref{le1-a},\ref{cross} and Equation (\ref{eqRw}), we have
\begin{eqnarray*}
R_{w}(\tau)&=&R_{\abu L^{\eta}(\bar{\abu})}(\frac{\tau-1}{4})+ R_{\bbu L^{\eta}(\bbu)}(\frac{\tau-1}{4})+ R_{L^{\eta}(\bar{\abu})\bbu}(\frac{\tau-1}{4})+R_{L^{\eta}(\bbu)\abu}(\frac{\tau+3}{4})\\
           &=&-R_{\abu}(\tau_1+\eta)+ R_{\bbu}(\tau_1+\eta)-R_{\abu\bbu}(\tau_1-\eta)
           +R_{\bbu \abu}(\tau_1+1-\eta).\\
           &=&-R_{\abu}(\tau_1+\eta)+ R_{\bbu}(\tau_1+\eta)-R_{L^{\frac{N+1}{2}}(\abu)\bbu}(\tau_1+\frac{N+1}{2}-\eta)
           +R_{\bbu L^{\frac{N+1}{2}}(\abu)}(\tau_1+\frac{N+1}{2}-\eta)\\
           &=&-R_{\abu}(\tau_1+\eta)+ R_{\bbu}(\tau_1+\eta)-R_{\abu \bbu}(\tau_1-\eta)
           +R_{\bbu\abu}(\tau_1+\frac{N+1}{2}-\eta).
\end{eqnarray*}

if $\tau_2=3$, then $\frac{\tau-1}{2}$ is odd, $\frac{\tau+1}{2}$ is even, from Lemmas \ref{abcd},\ref{le1-a},\ref{cross} and Equation (\ref{eqRw}),
\begin{eqnarray*}
R_{w}(\tau)\!\!\!&=&\!\!\!R_{\abu L^{\eta}(\bbu)}(\frac{\tau-3}{4})+ R_{\bbu L^{\eta}(\bar{\abu})}(\frac{\tau+1}{4})+ R_{L^{\eta}(\bar{\abu})\abu}(\frac{\tau+1}{4})+R_{L^{\eta}(\bbu)\bbu}(\frac{\tau+1}{4})\\
           &=&\!\!\!R_{\abu\bbu}(\frac{\tau-3}{4}+\eta)- R_{\bbu \abu}(\frac{\tau+1}{4}+\eta)-R_{\abu}(\frac{\tau+1}{4}-\eta)+R_{\bbu}(\frac{\tau+1}{4}-\eta)\\
        &=&\!\!\!R_{L^{\frac{N+1}{2}}(\abu)\bbu}(\tau_1+\frac{N+1}{2}+\eta)- R_{\bbu L^{\frac{N+1}{2}}(\abu)}(\tau_1+\frac{N+1}{2}+\eta)-R_{\abu}(\tau_1+1-\eta)+R_{\bbu}(\tau_1+1-\eta)\\
        &=&\!\!\!R_{\abu \bbu}(\tau_1+\eta)-R_{\bbu \abu}(\tau_1+\eta)-R_{\abu}(\tau_1+1-\eta)+R_{\bbu}(\tau_1+1-\eta).
\end{eqnarray*}
From Theorem \ref{main2}, we can induce three ideals to decrease the autocorrelation of the sequence $w$:

Ia: If $R_{\abu}(\tau_1+1-\eta)=R_{\bbu}(\tau_1+1-\eta)$,  we decrease the value  $\mid R_{\abu \bbu}(\tau_1+\eta)-R_{\bbu \abu}(\tau_1+\eta)\mid$,

Ib: If $ R_{\abu \bbu}(\tau_1+\eta)=R_{\bbu \abu}(\tau_1+\eta)$, we decrease the value  $\mid R_{\abu}(\tau_1+1-\eta)-R_{\bbu}(\tau_1+1-\eta)\mid$.

Ic: Otherwise, we have to decrease the values of  $\mid R_{\abu \bbu}(\tau_1+\eta)-R_{\bbu \abu}(\tau_1+\eta)\mid$ and  $\mid R_{\abu \bbu}(\tau_1+\eta)-R_{\bbu \abu}(\tau_1+\eta)\mid$ analogously.

It is obvious that all these ideals are based on the fact that $2R_{\abu}(\tau_1) +2R_{\bbu}(\tau_1)$  for $\tau_2 = 0$ possess low values appropriately.

For the generalized GMW construction $s$ and its modifications $s'$, if  $a=s,b=s'$, then, by the ideal Ib, we have the out-of-phase autocorrelation
\begin{eqnarray*}
&&R_{\wbu}(\tau) = \left\{
\begin{array}{llllll}
 -4 &\mbox{ if }  \tau_2=0 \mbox{ and }\tau_1\equiv0 \bmod 2^n+1,\\
 4 &\mbox{ if }   \tau_2=0 \mbox{ and }\tau_1\not\equiv0 \bmod 2^n+1,\\
 0 &\mbox{ if }   \tau_2=1 \mbox{ and }\tau_1+\eta\equiv0 \bmod 2^n+1,\\
 4 &\mbox{ if }   \tau_2=1 \mbox{ and }\tau_1+\eta\not\equiv0 \bmod 2^n+1,\\
 0 &\mbox{ if }   \tau_2=2,\\
 0 &\mbox{ if }   \tau_2=3 \mbox{ and }\tau_1+1-\eta\equiv0 \bmod 2^n+1,\\
 4 &\mbox{ if }   \tau_2=3 \mbox{ and }\tau_1+1-\eta\not\equiv0 \bmod 2^n+1;
\end{array}
\right.
\end{eqnarray*}
if  $a=s',b=s$, then, by the ideal Ib, we have the out-of-phase autocorrelation
\begin{eqnarray*}
&&R_{\wbu}(\tau) = \left\{
\begin{array}{llllll}
 -4 &\mbox{ if }  \tau_2=0 \mbox{ and }\tau_1\equiv0 \bmod 2^n+1,\\
 4 &\mbox{ if }   \tau_2=0 \mbox{ and }\tau_1\not\equiv0 \bmod 2^n+1,\\
 0 &\mbox{ if }   \tau_2=1 \mbox{ and }\tau_1-\eta\equiv0 \bmod 2^n+1,\\
 -4 &\mbox{ if }   \tau_2=1 \mbox{ and }\tau_1-\eta\not\equiv0 \bmod 2^n+1,\\
 0 &\mbox{ if }   \tau_2=2,\\
 0 &\mbox{ if }   \tau_2=3 \mbox{ and }\tau_1+\eta\equiv0 \bmod 2^n+1,\\
 -4 &\mbox{ if }   \tau_2=3 \mbox{ and }\tau_1+\eta\not\equiv0 \bmod 2^n+1,.
\end{array}
\right.
\end{eqnarray*}

For Legendre sequences $l(t)$ and $l_0(t)$ for $p\equiv3\bmod 4$, if  $a=l(t),b=l_0(t)$, then, by the ideal Ia, we have the out-of-phase autocorrelation of $w$

\begin{eqnarray*}
&&R_{\wbu}(\tau) = \left\{
\begin{array}{llllll}
 -4 &\mbox{ if }  \tau_2=0,\\
 -4 &\mbox{ if }   \tau_2=1 \mbox{ and }\tau_1-\eta\in QR_p,\\
 4 &\mbox{ if }  \tau_2=1 \mbox{ and }\tau_1-\eta\in NQR_p,\\
 0 &\mbox{ if }  \tau_2=1 \mbox{ and }\tau_1-\eta=0,\\
 0 &\mbox{ if }   \tau_2=2,\\
4 &\mbox{ if }   \tau_2=3 \mbox{ and }\tau_1+\eta\in QR_p,\\
 -4 &\mbox{ if }   \tau_2=3 \mbox{ and }\tau_1+\eta\in NQR_p.
\end{array}
\right.
\end{eqnarray*}

if  $a=l_0(t),b=l(t)$, then, by the ideal Ia, we have the out-of-phase autocorrelation of $w$

\begin{eqnarray*}
&&R_{\wbu}(\tau) = \left\{
\begin{array}{llllll}
 -4 &\mbox{ if }  \tau_2=0,\\
 4 &\mbox{ if }   \tau_2=1 \mbox{ and }\tau_1-\eta\in QR_p,\\
 -4 &\mbox{ if }  \tau_2=1 \mbox{ and }\tau_1-\eta\in NQR_p,\\
 0 &\mbox{ if }   \tau_2=2,\\
 -4 &\mbox{ if }   \tau_2=3 \mbox{ and }\tau_1+\eta\in QR_p,\\
 4 &\mbox{ if }   \tau_2=3 \mbox{ and }\tau_1+\eta\in NQR_p,\\
 0 &\mbox{ if }  \tau_2=3 \mbox{ and }\tau_1+1-\eta=0.
\end{array}
\right.
\end{eqnarray*}

For Legendre sequences $l(t)$ and $l_0(t)$ for $p\equiv1\bmod 4$, if  $a=l(t),b=l_0(t)$, then, by the ideal Ib, we have the out-of-phase autocorrelation of $w$

\begin{eqnarray*}
&&R_{\wbu}(\tau) = \left\{
\begin{array}{llllll}
 -4 &\mbox{ if }  \tau_2=0\\
 -4 &\mbox{ if }   \tau_2=1 \mbox{ and }\tau_1+\eta\in QR_p,\\
 4 &\mbox{ if }  \tau_2=1 \mbox{ and }\tau_1+\eta\in NQR_p,\\
 0 &\mbox{ if }  \tau_2=1 \mbox{ and }\tau_1-\eta=0,\\
 0 &\mbox{ if }   \tau_2=2,\\
-4 &\mbox{ if }   \tau_2=3 \mbox{ and }\tau_1+1-\eta\in QR_p,\\
 4 &\mbox{ if }   \tau_2=3 \mbox{ and }\tau_1+1-\eta\in NQR_p,\\
  0 &\mbox{ if }  \tau_2=3 \mbox{ and }\tau_1+1-\eta=0.
\end{array}
\right.
\end{eqnarray*}

if  $a=l_0(t),b=l(t)$, then, by the ideal Ib, we have the out-of-phase autocorrelation of $w$

\begin{eqnarray*}
&&R_{\wbu}(\tau) = \left\{
\begin{array}{llllll}
 -4 &\mbox{ if }  \tau_2=0,\\
 4 &\mbox{ if }   \tau_2=1 \mbox{ and }\tau_1+\eta\in QR_p,\\
 -4 &\mbox{ if }  \tau_2=1 \mbox{ and }\tau_1+\eta\in NQR_p,\\
 0 &\mbox{ if }  \tau_2=1 \mbox{ and }\tau_1-\eta=0,\\
 0 &\mbox{ if }   \tau_2=2,\\
 4 &\mbox{ if }   \tau_2=3 \mbox{ and }\tau_1+1-\eta\in QR_p,\\
 -4 &\mbox{ if }   \tau_2=3 \mbox{ and }\tau_1+1-\eta\in NQR_p,\\
  0 &\mbox{ if }  \tau_2=3 \mbox{ and }\tau_1+1-\eta=0.
\end{array}
\right.
\end{eqnarray*}

For the twin-prime sequence $t$ and its modifications $t'$, if  $a=t,b=t'$, then, by the ideal Ib, we have the out-of-phase autocorrelation
\begin{eqnarray*}
&&R_{\wbu}(\tau) = \left\{
\begin{array}{llllll}
 -4 &\mbox{ if }  \tau_2=0 \mbox{ and }\tau_1\equiv0 \bmod (p+2),\\
 4 &\mbox{ if }   \tau_2=0 \mbox{ and }\tau_1\not\equiv0 \bmod (p+2),\\
 0 &\mbox{ if }   \tau_2=1 \mbox{ and }\tau_1\equiv0 \bmod (p+2),\\
 4 &\mbox{ if }   \tau_2=1 \mbox{ and }\tau_1\not\equiv0 \bmod (p+2),\\
 0 &\mbox{ if }   \tau_2=2,\\
 0 &\mbox{ if }   \tau_2=3 \mbox{ and }\tau_1+1-\eta\equiv0 \equiv0 \bmod (p+2),\\
 4 &\mbox{ if }   \tau_2=3 \mbox{ and }\tau_1+1-\eta\not\equiv0 \bmod (p+2).
\end{array}
\right.
\end{eqnarray*}

 if  $a=t',b=t$, then, by the ideal Ib, we have the out-of-phase autocorrelation
\begin{eqnarray*}
&&R_{\wbu}(\tau) = \left\{
\begin{array}{llllll}
 -4 &\mbox{ if }  \tau_2=0 \mbox{ and }\tau_1\equiv0 \bmod (p+2),\\
 4 &\mbox{ if }   \tau_2=0 \mbox{ and }\tau_1\not\equiv0 \bmod (p+2),\\
 0 &\mbox{ if }   \tau_2=1 \mbox{ and }\tau_1\equiv0 \bmod (p+2),\\
 -4 &\mbox{ if }   \tau_2=1 \mbox{ and }\tau_1\not\equiv0 \bmod (p+2),\\
 0 &\mbox{ if }   \tau_2=2,\\
 0 &\mbox{ if }   \tau_2=3 \mbox{ and }\tau_1+1-\eta\equiv0 \equiv0 \bmod (p+2),\\
 -4 &\mbox{ if }   \tau_2=3 \mbox{ and }\tau_1+1-\eta\not\equiv0 \bmod (p+2).
\end{array}
\right.
\end{eqnarray*}

\end{document}